\documentclass{egpubl}
\usepackage{eg2026}
 
\ConferencePaper        
\CGFccbync

\usepackage[T1]{fontenc}
\usepackage{dfadobe}  

\usepackage{booktabs} 
\usepackage{soul}
\usepackage[ruled]{algorithm2e} 
\usepackage{multirow}
\usepackage{makecell}
\usepackage{dblfloatfix}
\usepackage{amsfonts}
\usepackage{amsmath}

\biberVersion
\BibtexOrBiblatex
\usepackage[backend=biber,bibstyle=EG,citestyle=alphabetic,backref=true]{biblatex} 
\electronicVersion
\PrintedOrElectronic

\ifpdf \usepackage[pdftex]{graphicx} \pdfcompresslevel=9
\else \usepackage[dvips]{graphicx} \fi

\usepackage{egweblnk} 

\usepackage[dvipsnames,svgnames]{xcolor}
\definecolor{olive}{rgb}{0.1,0.8,0.3}
\definecolor{mauve}{rgb}{0.48,0,0.72}
\definecolor{red}{rgb}{1,0,0}
\definecolor{junglegreen}{rgb}{0.113, 0.639, 0.5}

\newcommand{\titleshort}[1]{\textbf{Step2Motion}}

\SetAlFnt{\small}
\SetAlCapFnt{\small}
\SetAlCapNameFnt{\small}
\SetAlCapHSkip{0pt}

\title[Step2Motion: Locomotion Reconstruction from Pressure Sensing Insoles]{Step2Motion: Locomotion Reconstruction from \\ Pressure Sensing Insoles}

\author[J. L. Ponton et al.]
{\parbox{\textwidth}
    {\centering
    J. L. Ponton$^{1}$\orcid{0000-0001-6576-4528}
    E. Alvarado$^{2}$\orcid{0000-0003-3395-5674}
    L. G. Foo$^{2}$\orcid{0000-0002-6082-6002}
    N. Pelechano$^{1}$\orcid{0000-0002-1437-245X}
    C. Andujar$^{1}$\orcid{0000-0002-8480-4713}
    and M. Habermann$^{2}$\orcid{0000-0003-3899-7515}
    }
    \\
    {\parbox{\textwidth}
        {\centering
        $^1$Universitat Politècnica de Catalunya, Barcelona, Spain\\
        $^2$Max Planck Institute for Informatics, Saarbrücken, Germany
        }
    }
}

\begin{document}

\teaser{
 \includegraphics[width=1\linewidth,trim=5 25 5 25,clip]{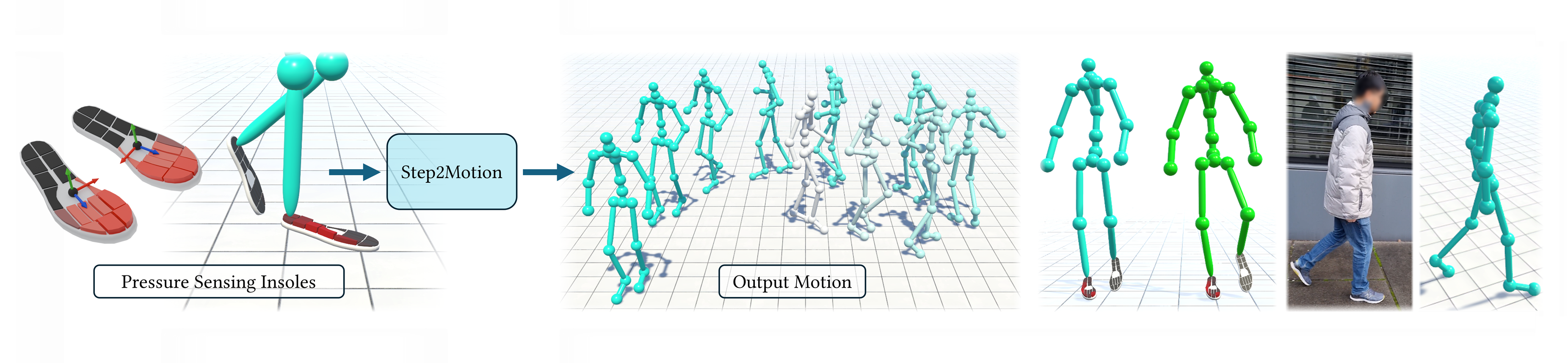}
 \centering
  \caption{\textbf{Locomotion reconstruction from pressure sensing insoles.} (Left) Our method achieves locomotion reconstruction using only data from insoles, each equipped with an IMU and 16 pressure sensors. (Middle) Representation of a user walking with the insole sensors, which are also able to capture the balanced pose when standing on one foot, among others. (Right) Reconstruction of \textit{in-the-wild} locomotion while the subject wears the insole sensors.}
\label{fig:teaser}
}

\maketitle

\begin{abstract}
Human motion is fundamentally driven by continuous physical interaction with the environment. 
Whether walking, running, or simply standing, the forces exchanged between our feet and the ground provide crucial insights for understanding and reconstructing human movement.
Recent advances in wearable insole devices offer a compelling solution for capturing these forces in diverse, real-world scenarios.
Sensor insoles pose no constraint on the users' motion (unlike mocap suits) and are unaffected by line-of-sight limitations (in contrast to optical systems). 
These qualities make sensor insoles an ideal choice for robust, unconstrained motion capture, particularly in outdoor environments.
Surprisingly, leveraging these devices with recent motion reconstruction methods remains largely unexplored.
Aiming to fill this gap, we present \titleshort{}, the first approach to reconstruct human locomotion from multi-modal insole sensors.
Our method utilizes pressure and inertial data\textemdash accelerations and angular rates\textemdash captured by the insoles to reconstruct human motion. 
We evaluate the effectiveness of our approach across a range of experiments to show its versatility for diverse locomotion styles, from simple ones like walking or jogging up to moving sideways, on tiptoes, slightly crouching, or dancing. 
The complete source code, trained model, data, and supplementary material used in this paper can be found at: \href{https://vcai.mpi-inf.mpg.de/projects/Step2Motion/}{https://vcai.mpi-inf.mpg.de/projects/Step2Motion/}



\end{abstract}

%
%
\section{Introduction}\label{sec:introduction}

Human motion reconstruction plays a crucial role in a wide range of fields, ranging from entertainment-related applications (games, VR) to those involving greater biomechanical complexity (sports, rehabilitation).
Consequently, there is a growing demand for high-quality, accurate motion capture systems. 
However, current technologies often present barriers that limit their widespread use. 
Optical (external or egocentric) and markerless systems, while accurate, are typically complex and expensive, require specialized equipment, and perform better in controlled environments, limiting their practicality in outdoor recordings~\cite{shiratori2011, rhodin2016, liu2021neural, li2023a, Shetty_2024_CVPR}. 
Although less constrained by the capture area, IMU-based systems require specialized suits or external attachments that restrict movement and require frequent calibration~\cite{Huang:2018, Yi:2021, Yi:2022:pip, Jiang:2022, yi:2023}. 
Applications such as sports analytics and injury prevention often require solutions where unconstrained movement and ease of use are vital. 
Ideally, such a system should be easy to set up and wear, capable of recording arbitrary motion outside a recording studio (e.g., mountain hiking), and robust enough to handle intense activity without sensor displacement or occlusions (e.g., rugby).
%
%
%
\par 
Contact dynamics could play an important role in the search for such new capture systems. 
Legged motion is fundamentally driven by the continuous action of the feet against the ground, producing, in return, a reaction force distributed over the contact area. 
Related factors such as center of pressure (CoP), distributed weight, or impact forces caused by the foot strike during the gait cycle are often enough to serve as an accurate descriptor of human movement \cite{Ren2008, Wannop2014, Shahabpoor2017}.
Traditionally, force plates have been used to measure dynamics during gait \cite{beckham2014force}.
More recently, wearable insole devices have proved to be an excellent way to measure gait dynamics in arbitrary environments \cite{reviewinsoles}, although their use has often been limited to motion analysis; both their hardware features and output data make them a potential candidate in reconstruction tasks, suitable for novel entertainment, VR or biomechanics.
%
%
\par 
This paper introduces \titleshort{}, a novel deep learning-based approach for reconstructing human locomotion and root motion using pressure and inertial measurements from insole sensors.
Our method leverages this information to condition a diffusion-based motion reconstruction model. 
We develop a new multi-head cross-attention mechanism to effectively incorporate the multi-modal nature of insole data, enabling the network to selectively attend to different sensor modalities based on the body part being reconstructed and capture the complex relationships between insole measurements and human movement.
%
%
\par 
To our knowledge, this is the first approach to achieve general locomotion reconstruction solely from insole sensor data.
While recent work such as Smart Insole \cite{han2024smartinsole} has demonstrated pose estimation from pressure data, our approach differs fundamentally in scope, modality, and hardware. Han et al. \cite{han2024smartinsole} rely on high-density research-grade sensor grids (>600 sensors per foot) to perform discrete activity classification and local pose estimation relative to the pelvis. In contrast, Step2Motion targets the distinct problem of continuous locomotion reconstruction using sparse, consumer-grade hardware (only 16 sensors per foot + IMU). We reconstruct the global root trajectory and synthesize temporally coherent animation via a diffusion framework, rather than performing per-frame keypoint regression.

We demonstrate the versatility of our proposed algorithm and hardware setup by reconstructing diverse locomotion styles, including walking, dancing, jogging, or tiptoeing.
Through extensive evaluation and analysis, we provide insight into our design choices and highlight the reconstruction effectiveness of our approach.
The main contributions of our paper can be summarized as follows:
\begin{itemize}
\item We propose \titleshort{}\textemdash the first method for general human locomotion reconstruction from insole sensors. Our system accurately reconstructs lower-body motion while synthesizing plausible upper-body movements that naturally align with the reconstructed motion.
\item We introduce a displacement predictor network that effectively predicts root motion displacement from insole sensors.
\item Along with a comprehensive set of evaluations, we recorded a new motion capture dataset paired with insole readings, which we have made publicly available under the following link: \href{https://vcai.mpi-inf.mpg.de/projects/Step2Motion/}{https://vcai.mpi-inf.mpg.de/projects/Step2Motion/}
\end{itemize}
%
%
\section{Related Work}
\label{sec:related_work}



\subsection{Motion Capture from Body-worn Sensors}
\label{sec:rw:mocap_bodyworn}
Optical-based motion capture systems \cite{vicon, optitrack} achieve highly accurate motion reconstruction but are primarily used in indoor environments with multi-camera setups. In recent years, motion capture with body-worn sensors \cite{xsens} has gained popularity due to its versatility in unconstrained outdoor environments and robustness to occlusions and lighting conditions. Body-worn sensor approaches can be broadly classified by sensor type: Inertial Measurement Units (IMUs), outside-in trackers, and egocentric cameras.

\paragraph*{Inertial Measurement Units.}
IMU-based mocap achieves precise pose reconstruction with multiple IMUs. Efforts to enhance accessibility have reduced sensor counts, starting with six IMUs and offline optimization~\cite{Marcard:2017}. For real-time use, deep learning architectures have emerged~\cite{Huang:2018, Yi:2021}. However, persistent issues remain, like drift errors from integrating acceleration and angular rate.
Subsequent methods address drift reduction: Jiang et al. \cite{Jiang:2022} apply Transformers for temporal data, while Yi et al. \cite{Yi:2022:pip} use physics-based optimization, later refining non-inertial effects~\cite{yi:2024:pnp}. Armani et al. \cite{ultra_inertial_poser} estimate inter-sensor distances to mitigate drift and jitter. Additionally, monocular cameras paired with SLAM algorithms~\cite{guzov:2021, yi:2023, lee:2024} help locate subjects and minimize drift.
Unlike full-body IMU suits prone to looseness, our method employs just two IMUs embedded in the insoles. This design leverages the feet's stability as natural anchors, reducing sensor displacement.

\paragraph*{Outside-in Trackers.}
The rising demand for VR has driven the development of high-accuracy positional and rotational sensors. Early efforts, such as~\cite{Yang:2021:lobstr}, used recurrent neural networks for motion reconstruction with four 6DoF sensors. To avoid additional tracking devices, Ponton et al. \cite{ponton2022mmvr} proposed a motion-matching system for one or three sensors. Other works with three sensors explored deep learning methods, including transformers~\cite{Jiang:2022:AvatarPoser}, reinforcement learning~\cite{Winkler:2022:QuestSim}, diffusion models~\cite{avatar_grow_legs}, and vector quantization~\cite{codebook_matching:starke}. Recent research focuses on enhancing reconstruction accuracy using six sensors~\cite{ponton2023sparseposer}, variable sensor counts~\cite{ponton2024dragposer}, and integrating 6DoF data with IMUs~\cite{divatrack}.
Insoles operate seamlessly outdoors and are free from restrictions, unlike positional sensors that depend on external devices. Despite progress in built-in cameras, these systems remain constrained by limited capture space and bright outdoor scenes.

\paragraph*{Egocentric Cameras.}
Egocentric camera-based estimation eliminates the need for additional tracking devices. However, it faces other challenges, like occlusion and lighting variation. Early approaches~\cite{shiratori2011, rhodin2016} tackled these using body-worn and dual fisheye cameras, further refined by a single head-mounted fisheye camera~\cite{xu2019}. Advances in deep learning enhanced fisheye-based estimation~\cite{jiang2021, tome2019, wang2021, tome2023, wang2023}. Zhang et al. \cite{zhang2021} improved lens distortion handling through automatic calibration, while physics-based controllers added realism~\cite{yuan2018, yuan2019}. Recently, Li et al. \cite{li2023a} simplified motion capture with a two-stage estimation process, bypassing paired egocentric videos and motions.
Such cameras often struggle with lower-body accuracy due to visibility constraints. In contrast, our method ensures precise contact estimation and infers accurate locomotion predictions, making it resilient to occlusions and ideal for advancing physics-based motion techniques.

\subsection{Pressure Sensors for Character Animation}
\label{sec:rw:pressure}
Foot pressure data is crucial for understanding locomotion, as it reflects the forces exerted on the ground. Traditionally analyzed with force plates~\cite{han2023groundlink}, this data is paired with kinematic mocap to predict ground reaction forces (GRFs) and the center of pressure (CoP).
%
Such ideas extend to pose generation, allowing animators to draw foot pressure maps and retrieve the corresponding poses from a database~\cite{yin2003footsee} or use the exerted impact to alter the character's locomotion~\cite{Alvarado2022} and environment~\cite{Alvarado2024}. Grimm et al. \cite{Grimm2011} employed force plates in a mattress to classify patient poses using nearest-neighbor search based on basic pressure patterns. Additionally, joint data has been employed to predict foot pressure maps~\cite{scott2020image}.

Wireless insole sensors provide temporal pressure distribution data per foot and integrate seamlessly into regular footwear. Despite their potential, research in this area remains limited. Most studies use insole sensors for precise foot contact labeling. Mourot et al. \cite{mourot2022underpressure} introduced a method to correct foot sliding artifacts, pairing mocap with insole data to train a model estimating vertical ground reaction forces (vGRF) during locomotion. Addressing the inverse problem, Wu et al. \cite{soleposer} focused on predicting skiing poses from insole data.
However, due to the limited input, their upper-body reconstruction appears restricted, and root motion is not shown. While these works focus on a single type of motion, we prove how our method can generalize over different types and motion styles.
Han et al. \cite{han2024smartinsole} presented a CNN-based approach for estimating 3D poses using insoles with over 600 pressure sensors per foot, focusing on activity classification rather than motion reconstruction. Our approach uses publicly available insoles, combining cross-modality inputs (pressure and IMU) to produce smooth animation data.



\subsection{Diffusion-based Motion Synthesis}
\label{sec:rw:diffusion}
Diffusion models~\cite{diffusion_models} have recently been applied to human motion synthesis from text and audio inputs. Approaches include U-Nets~\cite{mofusion} and transformers~\cite{tevet2023mdm, flame:2023, motiondiffuse:2024} to capture temporal dependencies.
Subsequent research has significantly advanced motion synthesis and reconstruction using various conditioning signals. For synthesis, Mughal et al. \cite{convofusion} generated gestures from audio via latent diffusion, while Sun et al. \cite{lgtm:sun:2024} employed LLMs for part-specific motion from text, and Cohan et al. \cite{cohan:2024} developed a text-conditioned motion in-betweening framework. In reconstruction, Zhang et al. \cite{tedi:zhang:2024} achieved long-term motion by encoding temporality into denoising. Motion was also reconstructed from noisy RGB(-D) videos by Zhang et al. \cite{rohm:zhang:2024}, and from IMUs by Van Wouwe et al. \cite{wouwe:diffusionposer:2024}, though the latter optionally used insoles only for contact labeling.
Despite significant progress in diffusion-based motion generation, prior works have not examined conditioning these models on insole data features, like pressure, acceleration, and total force. Therefore, we introduce the first diffusion-based model that successfully integrates these diverse modalities, overcoming challenges posed by their unique characteristics.

\section{Method}
\label{sec:method}

\begin{figure*}[ht]
    \centering
    \includegraphics[width=1.0\linewidth]{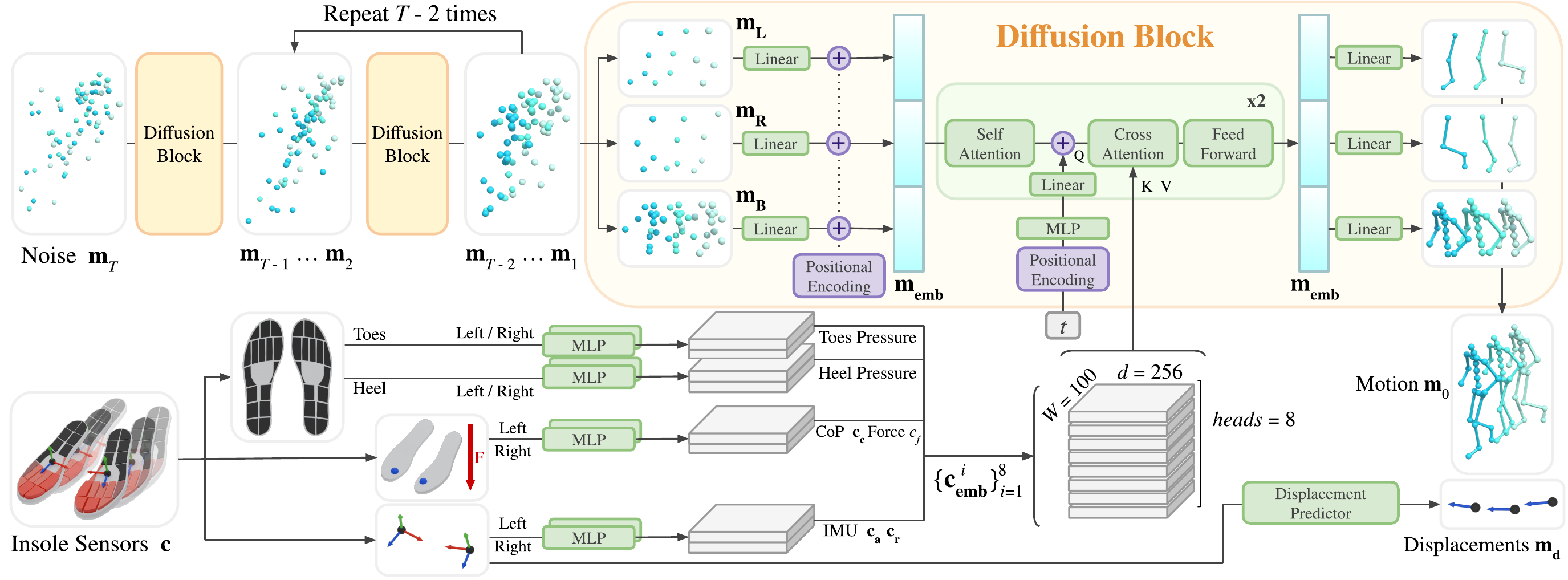}
    \caption{\textbf{Overview of the diffusion-based motion reconstruction process conditioned to insole sensor data.} The method starts with some unit gaussian noise $\mathbf{m}_T$ and returns the denoised motion sequence $\mathbf{m}_0$ after $T$ iterations. The diffusion block is executed each iteration: given a noisy input motion sample $\mathbf{m}_t$ at timestep $t$, the output is the denoised pose $\mathbf{m}_{t-1}$ at timestep $t-1$. The input is divided into three representations, $\mathbf{m_L}$, $\mathbf{m_R}$, $\mathbf{m_B}$, to facilitate part-wise attention within the Transformer network. The insole data $\mathbf{c}$ is partitioned into eight components to be used as heads for multi-head cross-attention. Sinusoidal positional encodings~\cite{transformers} are employed to encode motion temporality and the current diffusion timestep. An additional Transformer network predicts the displacements $\mathbf{m_d}$ of the corresponding motion $\mathbf{m}$ from the IMU readings of both feet.}
    \label{fig:pipeline:pose}
\end{figure*}

Our goal is to recover human locomotion solely from insole measurements (see Figure~\ref{fig:teaser}).
%
Such measurements are multimodal, featuring, for example, pressure and acceleration (data is explained in Section~\ref{sec:background}). 
%
Additionally, the limited sensing capabilities lead to inherent ambiguities; similar readings can correspond to the same pose. The multimodal nature and this inherent ambiguity present an interesting research question that we formalize in Section~\ref{sec:problem_definition}.
%
At the core, we solve this challenging task by introducing a transformer-based diffusion model with an attention mechanism to account for the data multimodality (as shown in~Figure~\ref{fig:pipeline:pose}).
We provide a high-level overview in Section~\ref{sec:overview}, followed by the detailed architecture in Section~\ref{sec:network_structure}.
%
%
\subsection{Insole Sensors}
\label{sec:background}

In this work, we use Moticon's OpenGo Sensor Insoles, which incorporate an Inertial Measurement Unit (IMU) in addition to 16 pressure sensors per foot. These are worn like regular shoe insoles and are available in different sizes. The raw data provided by each insole sensor includes:
\begin{itemize}
    \item \textbf{Pressure sensors, $\mathbf{c_p}\in\mathbb{R}^{16}$}: Each element represents the pressure applied to one of the 16 sensors distributed across the insole, measured in $1/4~\text{N}/\text{cm}^2$.
    \item \textbf{Linear acceleration, $\mathbf{c_a}\in\mathbb{R}^3$} is measured by the IMU in its local coordinate frame, in units of $g$. To facilitate the network's understanding of movement direction, we transform this information into a fixed world coordinate frame by integrating the angular rate data.
    \item \textbf{Angular rate, $\mathbf{c_r}\in\mathbb{R}^3$} is measured by the IMU in its local coordinate frame, in units of $\text{degree}/\text{s}$. 
    \item \textbf{Total force value, $c_f\in\mathbb{R}^1$} represents the ground reaction force magnitude measured in $\text{N}$.
    \item \textbf{Center of pressure, $\mathbf{c_c}\in\mathbb{R}^2 \cap  [-0.5, 0.5]$ } is a normalized CoP location as a percentage of the insole length and width.
\end{itemize}

\subsection{Problem Definition}
\label{sec:problem_definition}

Human motion can be represented as an ordered sequence of $W$ poses, $( \mathbf{p}^{(i)} )^W_{i=1}$, of a humanoid skeleton with $J$ joints. Each pose, $\mathbf{p}=(\mathbf{d}, \mathbf{j})$, comprises a world space displacement 3D vector $\mathbf{d} \in \mathbb{R}^3$ defining the root's 3D movement from the previous pose, and a set of relative joint positions $\mathbf{j} \in \mathbb{R}^{(J - 1) \times 3}$ represented as root-relative 3D vectors. The global position (i.e., the root position) in frame $i$ is the cumulative sum of displacements: $\sum^i_{f=0}\mathbf{d}^{(f)}$. %

In addition to poses, the insole readings $( \mathbf{c}^{(i)} )^W_{i=1}$ at the $i$-th frame is a 50-dimensional vector as follows:
\begin{equation}
\mathbf{c}^{(i)} = (\mathbf{c_p^L}, \mathbf{c_a^L}, \mathbf{c_r^L}, c_f^L, \mathbf{c_c^L}, \mathbf{c_p^R}, \mathbf{c_a^R}, \mathbf{c_r^R}, c_f^R, \mathbf{c_c^R}) \in \mathbb{R}^{50}
\end{equation} 
comprising 25 features per foot. Our objective is to synthesize a sequence of $W$ poses $( \mathbf{p}^{(i)} )^W_{i=1}$ from the corresponding insole readings $( \mathbf{c}^{(i)} )^W_{i=1}$.

\subsection{Overview}
\label{sec:overview}

Our approach to motion reconstruction from insole data is based on two primary components: a diffusion model for reconstructing poses (Figure~\ref{fig:pipeline:pose}). 
We employ a diffusion probabilistic model to synthesize poses due to its demonstrated ability to capture complex distributions and synthesize high-quality data. 
We operate on a sequence of poses, therefore capturing the temporal consistency of the motion.
To further enhance the synthesis, we incorporate the insole information through a carefully designed cross-attention mechanism that allows the network to focus on relevant sensor data based on the specific body part and the reconstructed motion. 
A separate Transformer network is used to regress the root displacements of the motion from the IMU data provided by the insole sensors. 
This two-stage pipeline allows for more effective processing of distinct feature types, preventing the displacement information from being ignored, a common issue even with input standardization.

\subsection{Network Structure}
\label{sec:network_structure}

\paragraph*{Diffusion Probabilistic Framework.}
We employ a diffusion probabilistic model framework~\cite{diffusion_models} consisting of a \emph{forward diffusion} and \emph{reverse diffusion} process. The \emph{forward diffusion} process is a Markovian chain that iteratively adds Gaussian noise to an initial sequence of poses, $\mathbf{m}^{(i)} = (\mathbf{j}^{(i)}, \ldots ,\mathbf{j}^{(i+W)}) \in \mathbb{R}^{W \times (J-1) \times 3}$, obtained by sliding a window of size $W$ over the complete pose sequence $( \mathbf{p}^{(i)} )^F_{i=1}$. This process transforms the input $\mathbf{m}_0$ into a sample from a standard Gaussian distribution $\mathbf{m}_T \sim \mathcal{N}(\mathbf{0}, \mathbf{I})$, after a number of timesteps $T$. Formally, the \emph{forward diffusion}, $q$, is defined as follows:
\begin{align}
    q(\mathbf{m}_t | \mathbf{m}_{t-1}) &= \mathcal{N}(\mathbf{m}_t; \sqrt{1-\beta_t}\mathbf{m}_{t-1}, \beta_t \mathbf{I}) \\
    q(\mathbf{m}_{1:T} | \mathbf{m}_0) &= \prod_{t=1}^T q(\mathbf{m}_t | \mathbf{m}_{t-1})
\end{align}
where $\{\beta_t \in (0,1) \}_{t=1}^T$ controls the variance schedule.

Conversely, the \emph{reverse diffusion} process, $p$, reconstructs a pose sequence from Gaussian noise. Starting from $\mathbf{m}_T \sim \mathcal{N}(\mathbf{0}, \mathbf{I})$, it progressively denoises the input over $T$ timesteps to recover the original pose sequence $\mathbf{m}_0$:
\begin{equation}
    p_{\theta}(\mathbf{m}_{0:T}) = p(\mathbf{m}_T) \prod_{t=1}^T p_{\theta}(\mathbf{m}_{t-1} | \mathbf{m}_t)
\end{equation}
where $p_{\theta}(\mathbf{m}_{t-1} | \mathbf{m}_t)$ is approximated utilizing a neural network $f_{\theta}(\mathbf{m}_{t-1} | \mathbf{m}_t, t,\mathbf{c})$ conditioned on the previous denoised sequence $\mathbf{m}_t$, the timestep $t$, the corresponding insole readings $\mathbf{c} \in \mathbb{R}^{W \times 50}$, and with parameters $\theta$. Following Tevet et al. \cite{tevet2023mdm} and Du et al. \cite{avatar_grow_legs}, we directly predict the denoised poses $\mathbf{m}_{t-1}$ at each step.

\paragraph*{Body-Partitioned Pose Encoding.}
Motivated by the success of Transformers~\cite{transformers} in motion synthesis~\cite{tevet2023mdm, convofusion, EMDM}, we use them as the core of $f_{\theta}$.
To leverage the inherent structure of the data, we decompose each pose into three body parts: left leg, right leg, and the remaining body.
This does not imply using independent Transformers. Instead, we generate a distinct vector embedding for each body part at every frame. These tokens are concatenated into a single sequence $\mathbf{m}_{emb}$. Consequently, the Transformer applies self-attention globally across both spatial (body parts) and temporal (frames) dimensions. This design enhances communication, enabling the network to be highly selective; for example, the left leg token at frame $i$ can attend to the right leg token at frame $i-5$.
This decomposition also allows the Transformer's attention mechanism to focus on relevant body parts based on the insole readings. Since each insole primarily influences the corresponding leg, this separation guides the network to prioritize pertinent information. 

Specifically, we partition the pose sequence $\mathbf{m} \in \mathbb{R}^{W \times (J-1) \times 3}$ into (assuming four joints per leg) $\mathbf{m_L} \in \mathbb{R}^{W \times 4 \times 3}$, $\mathbf{m_R} \in \mathbb{R}^{W \times 4 \times 3}$, and $\mathbf{m_B} \in \mathbb{R}^{W \times (J-1-8) \times 3} )$, representing the left leg, right leg, and the rest of the body, respectively. Each part is then projected to the Transformer's embedding dimension $d$ and augmented with sinusoidal positional encodings~\cite{transformers}. Time embeddings are generated by passing the sinusoidal positional embedding through a two-layer MLP. The final input to the Transformer layers, $\mathbf{m}_{emb} \in \mathbb{R}^{3W \times d}$, is formed by concatenating these embedded representations along the temporal dimension. 

We employ two Transformer layers with specific modifications to incorporate the insole conditioning effectively. First, a standard self-attention block processes the embedded pose sequence $\mathbf{m}_{emb}$, serving as the query, key, and value. Next, we encode the diffusion timestep $t$ using sinusoidal positional embeddings passed through a two-layer MLP.  This time embedding is then transformed by a linear layer specific to each decoder layer. By adding these timestep embeddings to the output of the self-attention block, we explicitly inform the network about the current stage of the diffusion process.

\paragraph*{Insole Multi-head Cross-Attention.}
Similarly to our body-partitioned pose encoding, we also separately process the cross-modality insole information to allow the attention mechanism to focus on specific sensor data based on the motion. 

Given the insole data, $\mathbf{c} \in \mathbb{R}^{W \times 25 \times 2}$, for the pose sequence, we generate four components per foot.
We generate the first two components by partitioning the pressure sensor readings $\mathbf{c_p}$ into two regions per foot: toes and heel (as visualized in Figure~\ref{fig:pipeline:pose}).
Next, we aggregate the IMU data ($\mathbf{c_a}$ and $\mathbf{c_r}$) into a single component. Lastly, we combine the total force $c_f$ and the center of pressure $\mathbf{c_c}$ into the fourth component.
Each of these four components per foot is then processed by a separate three-layer MLP to produce insole embeddings, $\{ \mathbf{c}^i_{emb} \in \mathbb{R}^{W \times d} \}_{i=1}^8$.

To efficiently integrate the insole embeddings into the motion reconstruction process, we employ a multi-head cross-attention block that treats each of the eight insole components (four per foot) as a separate attention head (see Figure~\ref{fig:pipeline:pose}). This allows the network to selectively attend to different aspects of the insole data based on the reconstructed motion without increasing the required computation by adding additional cross-attention blocks or increasing the attention matrix size.
%
More formally, our multi-head cross-attention mechanism can be expressed as:
\begin{align}
    \text{MultiHead}(\mathbf{m}_{emb} , \{ \mathbf{c}^i_{emb} \}_{i=1}^8) = \text{Concat}(\{ \mathbf{H}^i \}_{i=1}^8)\mathbf{W_O} \\
    \mathbf{H}^i = \text{Attention}(\mathbf{m}_{emb}            \mathbf{W}^i_{\mathbf{Q}} , 
                                     \mathbf{c}^i_{emb}\mathbf{W}^i_{\mathbf{K}} ,
                                     \mathbf{c}^i_{emb}\mathbf{W}^i_{\mathbf{V}})
\end{align}
where $\mathbf{W}^i_{\mathbf{Q}} \in \mathbb{R}^{d \times d/8}$, $\mathbf{W}^i_{\mathbf{K}} \in \mathbb{R}^{d \times d/8}$, $\mathbf{W}^i_{\mathbf{V}} \in \mathbb{R}^{d \times d/8}$, and $\mathbf{W_O} \in \mathbb{R}^{d\times d}$ are learnable parameter matrices, and $\mathbf{m}_{emb}$ is the output of the self-attention block with the added timestep embedding.

Following the cross-attention, a standard feed-forward block (two linear layers with activation and dropout functions) further processes the representation. Finally, after the two Transformer layers, the output $\mathbf{m}_{emb}$ is projected back to the original pose space, yielding the reconstructed 3D pose $\mathbf{\hat{m}}$.
The \emph{reverse diffusion} process is trained with the mean absolute error (MAE) loss between the reconstructed $\mathbf{\hat{m}}$ and the original pose sequence $\mathbf{m}$.

By designing the attention mechanism to map specific sensor groups to distinct heads, we introduce a strong inductive bias that aids interpretability and convergence. This separation allows the network to dynamically shift focus based on the gait phase. For example, during a heel strike, the network can attend heavily to the heel head to resolve contact, whereas during the swing phase, it can shift attention to the IMU head to track leg orientation. This prevents the high-dimensionality of the combined signal from diluting the critical cues provided by specific sensors.

\paragraph*{Displacement Predictor.}
In addition to predicting poses, we independently estimate the displacements, $\mathbf{m_d}^{(i)} = (\mathbf{d}^{(i)}, \ldots, \mathbf{d}^{(i+W)})$, for the sequence of poses $\mathbf{m}^{(i)}$. Instead of using a diffusion process, we directly regress the displacements from the IMU data using two standard Transformer layers with self-attention and feed-forward blocks. This choice is motivated by the observation that a given sequence of IMU readings typically corresponds to a specific displacement pattern, making this a suitable regression task. However, for pose reconstruction, where a one-to-many mapping may exist between pressure readings and corresponding poses, the diffusion process proves more suitable for modeling this inherent ambiguity.

We found that including both IMU and pressure data for displacement prediction often leads to overfitting the training data (see Section~\ref{sec:accuracy:displacement}). This happens because the displacement predictor network tends to memorize specific pressure patterns, lowering the performance for unseen motion. Consequently, we opted to utilize only IMU data to regress root displacements. However, larger training datasets might benefit from using both IMU and pressure data.

The IMU data, $(\mathbf{c_a^L}, \mathbf{c_r^L}, \mathbf{c_a^R}, \mathbf{c_r^R}) \in \mathbb{R}^{12}$, is first processed by a two-layer MLP to embed it into the Transformer's embedding dimension. Sinusoidal positional encodings are then added, and a final linear layer projects the output back to the displacement space. This approach allows us to leverage the temporal information within the IMU data to predict the root motion of the skeleton accurately.
The displacement predictor is trained with the mean squared error (MSE) loss between the reconstructed displacements $\mathbf{\hat{m_d}}$ and the original displacements $\mathbf{m_d}$. 

To improve accuracy and account for accumulated errors over time, we incorporate the cumulative sum of displacements within the loss function. The final loss for the displacement predictor is defined as
\begin{equation}
\label{eq:loss:displacement}
\mathcal{L} = \text{MSE}(\mathbf{m_d} , \mathbf{\hat{m_d}}) + \frac{\lambda}{W} \sum_{k=1}^{W} \text{MSE}\left(\sum_{i=1}^{k} \mathbf{d}^{(i)}, \sum_{i=1}^{k} \mathbf{\hat{d}}^{(i)}\right),
\end{equation}
where $\mathbf{\hat{d}}^{(i)}$ is the predicted displacement and $\lambda = 0.001$ in our experiments. 
This additional term ensures that the model is penalized for accumulating errors at any point within the temporal window, leading to more accurate displacement predictions throughout the motion sequence.

\subsection{Training and Implementation Details}
\label{sec:implementation}

We implemented \titleshort{} in PyTorch and used the Adam optimizer~\cite{kingma2017adam} for training. The experiments were conducted on a single NVIDIA GeForce RTX 4090 GPU. Both networks were trained with a batch size of 256 and a learning rate of $10^{-3}$. The pose diffusion model was trained for 500 epochs, while the displacement predictor was trained for 200 epochs. For the diffusion model, we followed the standard training procedure by sampling a random diffusion timestep $t$ from a uniform distribution and training the network to predict the denoised motion at the previous timestep, $\mathbf{m}_{t-1}$. Data augmentation was performed by randomly rotating the motion along the world's vertical axis.

We set the Transformer's embedding dimension to 256 and the feedforward block's dimension to 512. The diffusion process utilized 200 diffusion steps ($T$) with variances $\beta_t$ increasing linearly from 0.0001 to 0.02. We employed a window size $W$ of 100 poses, sampled at $30\,\text{Hz}$. We use GeLU activation functions similar to previous work \cite{convofusion}.

To reconstruct temporally coherent motion of arbitrary lengths, we implement an autoregressive approach inspired by the diffusion inpainting technique~\cite{diffusion_inpainting, convofusion}.
\section{Experiments and Evaluation}
\label{sec:evaluation}

 \begin{figure*}
     \centering
     \includegraphics[width=1\linewidth]{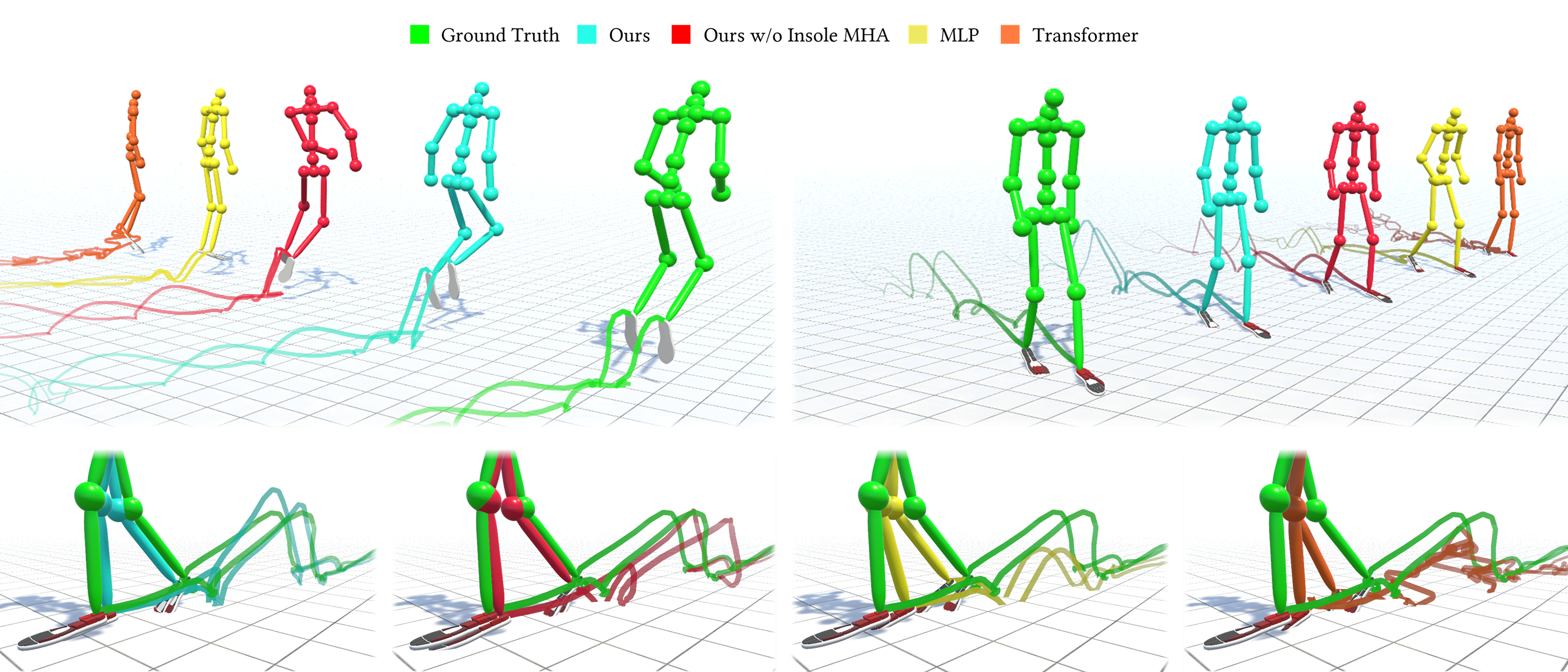}
     \caption{\textbf{Comparison on a jump followed by walking motion sequence.} Root motion is aligned to the ground truth (with an offset for visualization) to highlight pose differences. Our full method (cyan) accurately captures the jump trajectory and overall motion, while the \textbf{Transformer} baseline (orange) exhibits significant jitter and the \textbf{MLP} baseline (yellow) produces overly smooth motion. Removing the insole multi-head cross-attention (red) leads to a degradation in accuracy. The bottom row provides close-up views.
     }
     \label{fig:baseline:jump}
 \end{figure*}

 \begin{figure*}
    \centering
    \includegraphics[width=1\linewidth, trim=0 0 0 0,clip]{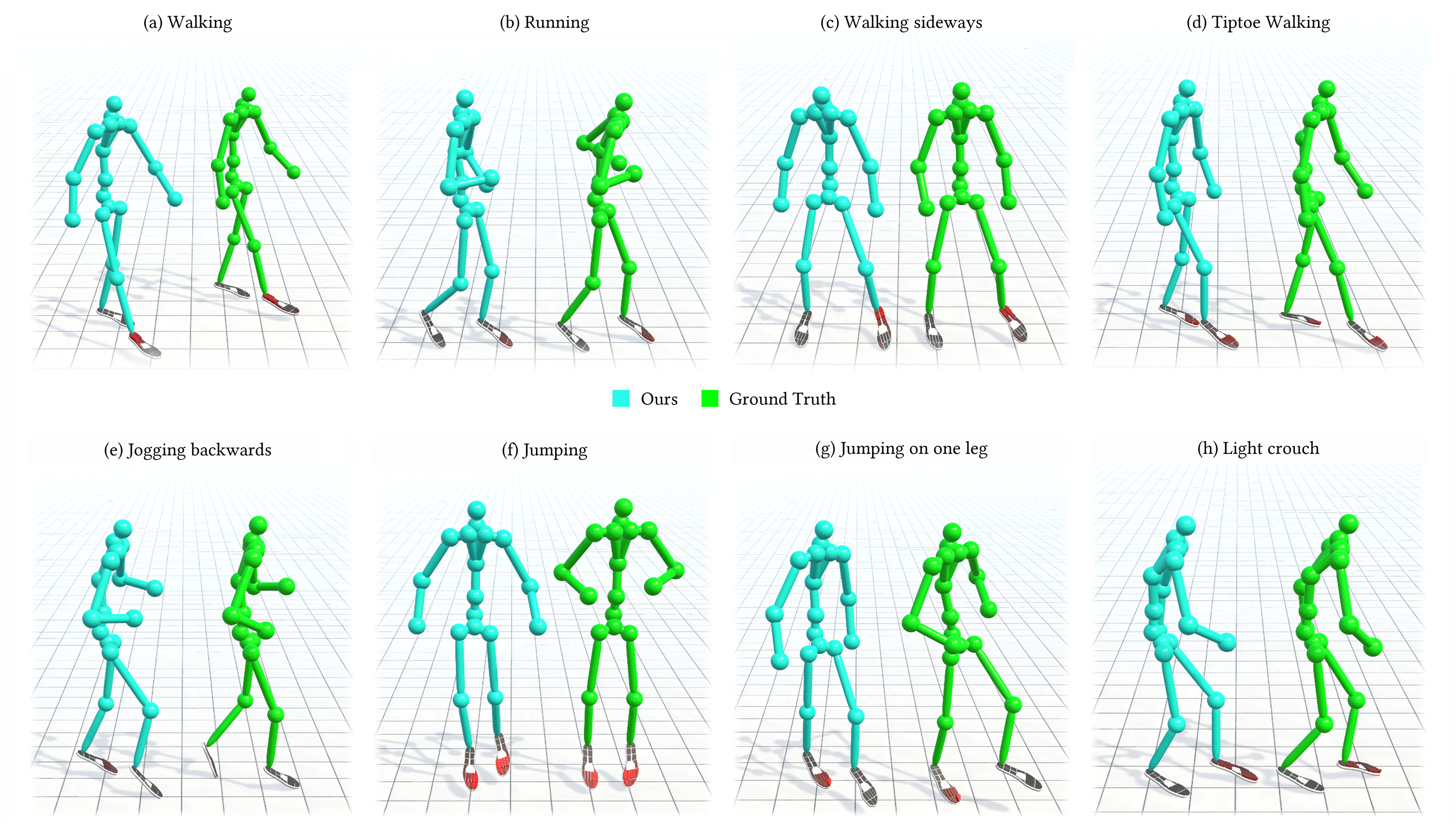}
    \caption{
    \textbf{Reconstructing locomotion styles.} Our method allows the reconstruction of different motion styles only using insole data, from walking and running to crouching, walking sideways, or moving on tiptoes.
    }
    \label{fig:experiments:diverse_motion}
\end{figure*}

In this section, we comprehensively evaluate our method on a publicly available dataset, employing both quantitative and qualitative assessments. We compare \titleshort{} with established baseline models and perform ablation studies to provide insight into our design choices. Pose accuracy is analyzed in Section~\ref{sec:accuracy:pose} and displacement prediction in Section~\ref{sec:accuracy:displacement}. Next, we also show the system's capabilities in diverse settings in Section~\ref{sec:experiments}. Finally, we analyze the use of pressure sensors and IMUs in Section~\ref{sec:eval:insole}.

\subsection{Datasets}
We use two datasets to train and evaluate \titleshort{}, UnderPressure~\cite{mourot2022underpressure}, and our own recorded locomotion dataset focusing on motion diversity, which we made publicly available under the link: \href{https://vcai.mpi-inf.mpg.de/projects/Step2Motion/}{https://vcai.mpi-inf.mpg.de/projects/Step2Motion/}

\paragraph*{UnderPressure Dataset.}
Public dataset comprising recordings of nine subjects performing various locomotion tasks such as walking, jogging, or jumping. 
It provides approximately 5 hours of motion capture data paired with insole readings. However, this dataset has very limited motion variety (normal pace walking and jogging), and subjects were instructed to perform actions in a very specific manner. 

\paragraph*{Step2Motion Dataset.}
To assess the performance in scenarios of greater motion diversity, we recorded our dataset. It includes markerless motion capture data paired with insole readings for $8$ subjects performing diverse locomotion activities.
Participants performed a continuous 25-minute sequence where the type of movement changed approximately every minute based on auditory instructions (e.g., walk, jog, squat). Subjects were explicitly instructed to perform the motion naturally without rigid constraints to ensure diversity.
For each type of movement, the participant performs that activity, then transitions between that activity and walking interchangeably, and finally performs the activity stationary for a specified time.
Total duration across subjects encompasses approximately $3.6$ hours of data.

All motion sequences used in the evaluation and depicted in the figures were not included during training. Furthermore, the testing sets for UnderPressure~\cite{mourot2022underpressure} and our dataset exclusively utilize motion data from subjects not included in the training sets. For instance, in the UnderPressure dataset, we test on Subject 4, who was intentionally selected due to significantly different body proportions compared to the training set (S1–S9). Subject 4 weighs 65\,kg with a height of 167\,cm, whereas all training subjects average significantly higher weight and height (e.g., S1 is 91\,kg/175\,cm; S7 is 88\,kg/184\,cm).

\subsection{Metrics}
We use four metrics in our evaluations, three of which to measure pose quality: \emph{Mean Per Joint Positional Error (MPJPE)}, which calculates the mean Euclidean distance between corresponding joints in centimeters; $\emph{MPJPE}_{\text{Legs}}$, identical to the previous one but considering the four joints of each leg only; and \emph{Mean Per Joint Velocity Error} ($\emph{MPJVE}_{\text{Legs}}$), which calculates the mean velocity error across the legs joints in centimeters per second. To analyze pose quality, the root is aligned with the ground truth. The fourth metric, \emph{Mean Root Positional Error (MRPE)}, measures the accuracy of the displacement predictor with the mean Euclidean distance error of the root joint in centimeters.

\subsection{Analysis of Pose Accuracy}
\label{sec:accuracy:pose}
 
 \begin{table*}[ht]
     \caption{\textbf{Pose accuracy evaluation} of our method, with and without insole multi-head cross-attention, compared with common deep learning architectures: \textbf{MLP} and \textbf{Transformer} (see Section~\ref{sec:accuracy:pose}).
     The standard deviation is shown in parentheses. Errors are reported in cm and cm/s.}
     \centering
     \begin{tabular}{lcccccc}
     \toprule
         \multirow{2.5}{*}{Method} & \multicolumn{3}{c}{UnderPressure Dataset~\cite{mourot2022underpressure}} & \multicolumn{3}{c}{Step2Motion Dataset} \\
     \cmidrule(lr){2-4} \cmidrule(l){5-7}
         & \emph{MPJPE} $\downarrow$ & $\emph{MPJPE}_{\text{Legs}}$ $\downarrow$ & $\emph{MPJVE}_{\text{Legs}}$ $\downarrow$ & \emph{MPJPE} $\downarrow$ & $\emph{MPJPE}_{\text{Legs}}$ $\downarrow$ & $\emph{MPJVE}_{\text{Legs}}$ $\downarrow$ \\
     \midrule
         MLP & 7.7(7.3) & 9.9(9.5) & 65.2(73.3) & 11.9(11.5) & 13.7(12.7) & 52.3(61.8) \\
         Transformer & 10.7(9.1) & 13.5(10.9) & 131.1(134.2) & 14.2(11.9) & 15.0(11.7) & 140.2(142.7) \\
         Ours w/o Insole MHA & 7.4(7.5) & 7.2(8.9) & 26.4(30.0) & 12.2(11.7) & 13.0(11.9) & 52.5(61.6) \\
         Ours & \textbf{7.2(7.1)} & \textbf{6.5(8.2)} & \textbf{26.1(29.2)} & \textbf{11.4(11.7)} & \textbf{12.3(11.9)} & \textbf{50.4(62.0)} \\
     \bottomrule
     \end{tabular}
     \label{tab:accuracy:pose}
 \end{table*}

We first performed an ablation study to evaluate the performance of \titleshort{} in reconstructing general full-body locomotion from insole sensor data. 
For this evaluation, acceleration is transformed from local to world space using ground truth rotational data, rather than through integration of angular rate. This isolates the performance of our method from the inherent challenges of IMU integration. 
The quantitative results of both datasets are summarized in Table~\ref{tab:accuracy:pose}, and qualitative comparisons are provided in Figure~\ref{fig:baseline:jump}.
Note that we separately train our system in the UnderPressure~\cite{mourot2022underpressure} database and our dataset.

We benchmark our method against two established architectures\textemdash \textbf{MLP} and \textbf{Transformer}\textemdash for regression problems. The Transformer model is a stack of Transformer encoders. 
These two baselines take a window of insole data as input and directly predict the corresponding window of poses. This comparison allows us to assess the benefits of using a diffusion-based approach over direct regression and to validate the architectural modifications we introduce for effectively leveraging insole data. 
Additionally, we perform an ablation study by removing the insole multi-head cross-attention and replacing it with a standard cross-attention module. This setting (\textbf{Ours w/o Insole MHA}) effectively evaluates a standard Transformer architecture within our diffusion framework.

To the best of our knowledge, this is the first approach for general locomotion reconstruction from insole sensors. The closest work is SolePoser~\cite{soleposer}, which focuses on the specific activity of skiing and does not synthesize arm and root motion. Due to their focus on skiing and the unavailability of their code, we do not include it in our comparison.

\begin{figure}
 \centering
 \includegraphics[width=0.95\linewidth]{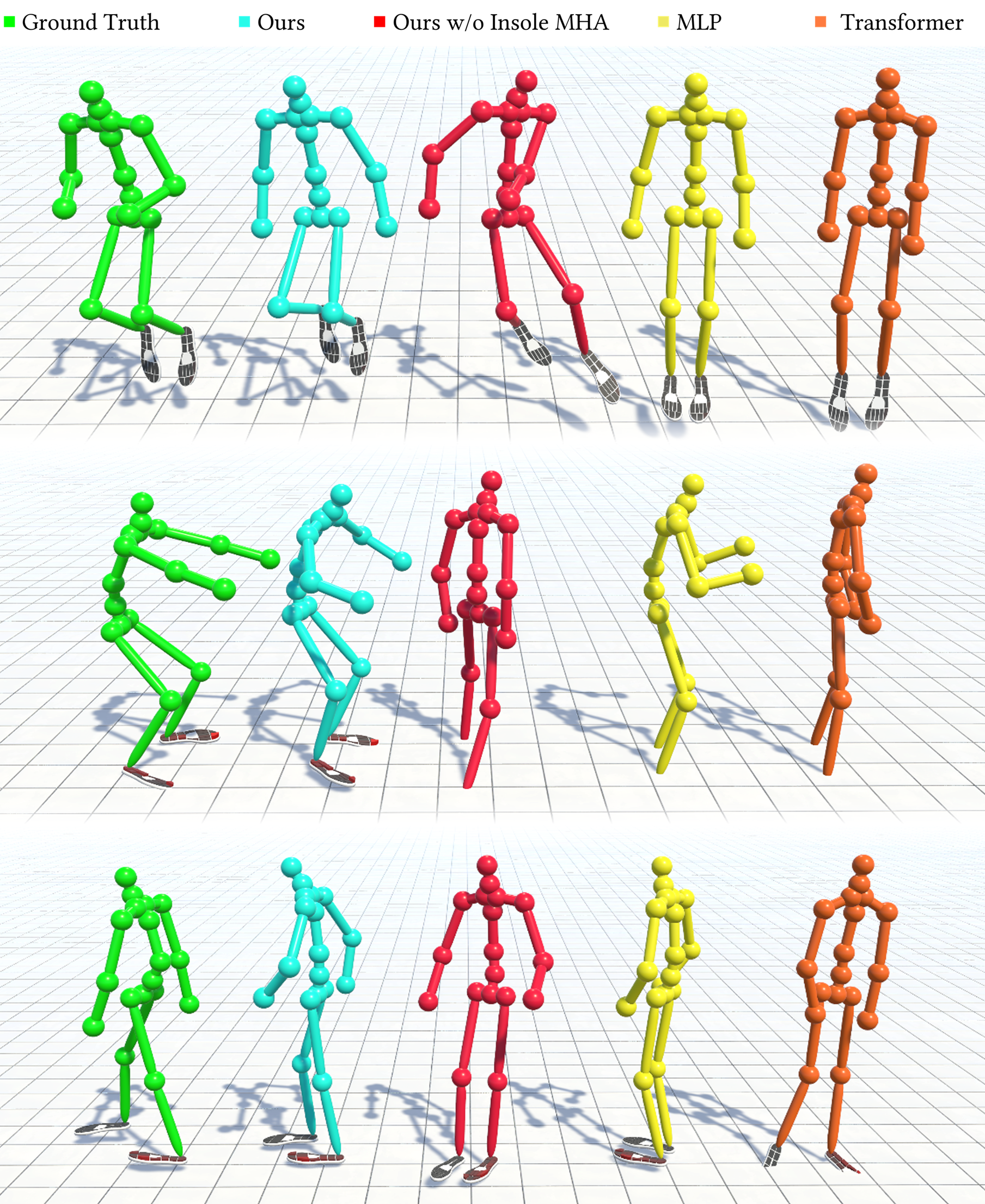}
 \caption{\textbf{Qualitative comparison on various motion sequences.} Root motion is aligned to the ground truth (with a slight offset for visualization purposes) to highlight pose quality differences. These animations were not used for training. The top image shows the same jump as Figure~\ref{fig:baseline:jump} but captured at the highest point. The middle and bottom images show squatting and walking motion, respectively. Our full method (cyan) accurately reconstructs various motions, including in-place movements like squatting. Removing the Insole MHA (red) leads to errors, particularly in sequences where distinct sensor information is crucial, such as the squat (relying on pressure) and walking (relying on IMU). The \textbf{MLP} (yellow) and \textbf{Transformer} (orange) baselines exhibit limitations in capturing these motions and overall, smoothed-out results.}
 \label{fig:baseline:details}
\end{figure}

Figure~\ref{fig:baseline:jump} visualizes the foot joint trajectories for each baseline, our method, and the mocap ground truth, for a sequence involving a jump followed by walking. Note that the root position is fixed to the ground truth to focus on pose quality. The \textbf{Transformer} baseline struggles to capture temporal consistency, producing highly jittery results with unpredictable trajectories, likely due to the limitations of the attention mechanism without the iterative refinement of a diffusion process. The \textbf{MLP} baseline reconstructs mostly correct poses but tends towards overly smooth and averaged motions (see Figure~\ref{fig:baseline:details}-top, which shows the same jump but in the peak height), as evident in the reduced amplitude of the jump trajectory. In contrast, our method accurately preserves the jump trajectory, particularly when using our full approach, with curves closely resembling the ground truth. For more qualitative ablations, we also refer to our supplemental video.

These observations are supported by the quantitative results in Table~\ref{tab:accuracy:pose}. The \textbf{Transformer} baseline consistently performs the worst across all metrics. The \textbf{MLP} baseline exhibits relatively low \emph{MPJPE} due to its tendency to produce average poses, but performs poorly when specifically analyzing leg movement (positions and velocities), indicating that high-frequency details are lost. Incorporating the insole multi-head cross-attention mechanism consistently improves all metrics, particularly the positional accuracy of the legs. 

The insole multi-head cross-attention mechanism is particularly crucial, as each modality should be analyzed differently depending on the action, rather than treating all sensor readings uniformly. In general, while walking, the IMU provides strong motion cues. However, during stationary actions such as squatting, the pressure distribution becomes the primary source of information. Figure~\ref{fig:baseline:details} highlights these scenarios. In the middle image, the full model correctly captures the squatting motion. This highlights the model's ability to prioritize pressure data from the insole when IMU readings are minimal.
Additionally, the image at the bottom shows that when the model cannot separately pay attention to the readings of the inertial sensors, it struggles to understand the direction of movement, which is determined primarily by the IMUs and not the pressure sensors. This leads to wrong global rotations, similar to a key limitation observed in the base \textbf{Transformer} model, which similarly uses the standard attention mechanism. This highlights the effectiveness of our approach in capturing fine-grained movements and reconstructing more realistic and detailed locomotion.

\subsection{Analysis of Root Displacement Accuracy}
\label{sec:accuracy:displacement}

In this section, we evaluate the performance of \titleshort{} in predicting the root position from per-frame displacements. Following the methodology in Section~\ref{sec:accuracy:pose}, we train the model separately on the UnderPressure~\cite{mourot2022underpressure} dataset and our dataset. Quantitative results are presented in Table~\ref{tab:accuracy:displacement}. Note that our dataset presents larger displacement errors due to the increased motion diversity and the length of the test animation clip (around 16\,min).

We propose different baselines and ablation experiments for displacement prediction compared to those used in Section~\ref{sec:accuracy:pose} for pose reconstruction, as we employ a separate network for this task.
We estimate per-frame displacement using \textbf{Double Integration} of the IMU acceleration and averaging the results (providing a non-data-driven approximation of root motion) as well as using an \textbf{MLP}.
The \textbf{Combined} baseline predicts the displacement within the diffusion process by adding a displacement term to the body-partitioned pose encoding, eliminating the need for a separate displacement predictor.
Finally, we compare these baselines with our method under different settings: using \textbf{Only Pressure} to predict the root displacement (i.e., the 16 pressure sensors, the center of pressure, and the total force per foot); \textbf{w/o Cumsum Loss}, by removing the second loss term in Equation~\ref{eq:loss:displacement} and making our system less aware of accumulated errors during training; and using both, \textbf{Pressure + IMU}, for the displacement predictor.
Differently from Section~\ref{sec:accuracy:pose}, the \textbf{Transformer} baseline is excluded from the displacement evaluation, as our displacement predictor uses a standard transformer encoder.

Analysis of Table~\ref{tab:accuracy:displacement} reveals that the double integration of acceleration signals yields poor results. This is likely due to the insole accelerometers producing inaccurate readings during foot-ground contact (due to sudden contact with the ground), leading to incorrect root motion estimation. A higher sampling frequency may help alleviate this issue, as we currently sample accelerations at $30\,\text{Hz}$.

\begin{figure}
     \centering
     \includegraphics[width=1.0\linewidth,trim=0 5 0 0,clip]{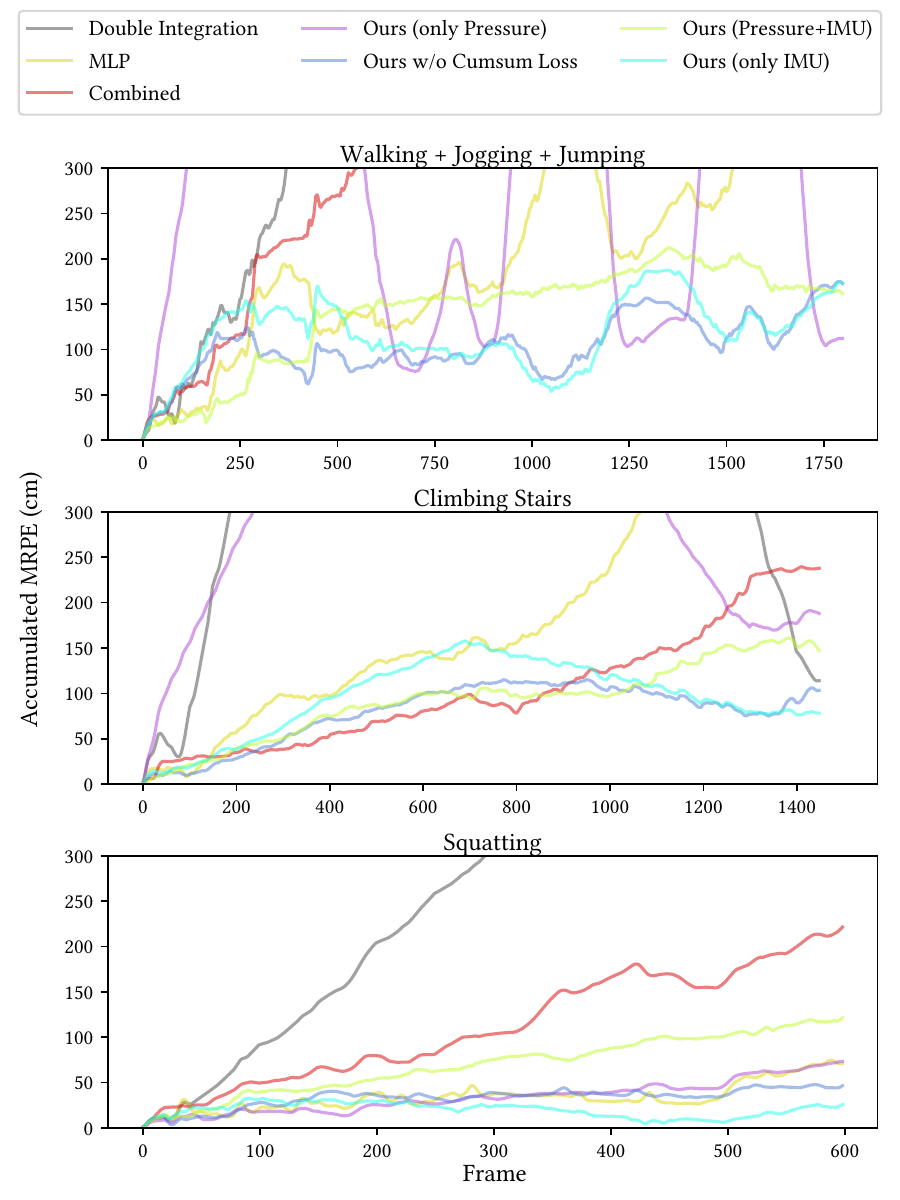}
     \caption{\textbf{Accumulated root positional error over time.} The figure demonstrates the limitations of double integration, pressure-only inputs, MLP architectures, and integrating displacement prediction into the diffusion process. Our full method, with or without pressure data, consistently shows the lowest error.}
     \label{fig:displacement:error}
 \end{figure}

Furthermore, providing only pressure information proves to be insufficient for an accurate prediction. Acceleration and angular rate from IMUs are strong predictors of root motion, as demonstrated in Figure~\ref{fig:displacement:error}, where pressure-only predictions perform well only in scenarios with minimal horizontal root motion (e.g., squatting). 

\begin{table}[htb] 
\caption{\textbf{Root displacement accuracy evaluation}. \textit{Mean Root Positional Error (MRPE)} is reported in meters, with standard deviation in parentheses.} 
\centering 
\begin{tabular}{lcc} 
\toprule 
\multirow{3}{*}{Method}  & UnderPressure     & Step2Motion  \\ 
                         &  Dataset $\downarrow$ &  Dataset $\downarrow$ \\ 
                         &  \cite{mourot2022underpressure} &   \\ 
\midrule 
Double Integration & 6.56(6.40) & 102.5(76.9) \\
MLP & 1.97(1.19) & 12.6(8.15) \\
Combined & 3.14(2.04) & 14.7(9.45) \\
Ours (only Pressure) & 4.57(2.65) & 10.1(7.06) \\
Ours w/o Cumsum Loss & 1.07(0.69) & 26.7(13.6) \\
Ours (Pressure+IMU) & \textbf{0.94(0.56)} & 10.1(6.26) \\
Ours (only IMU) & 0.97(0.54) & \textbf{6.41(4.03)} \\ 
\bottomrule 
\end{tabular} 
\label{tab:accuracy:displacement} 
\end{table}

Alternative architectures, such as MLPs or integrating displacement prediction into the diffusion process, also prove ineffective. The former likely struggles to capture temporal dependencies, while the latter may be hindered by the difficulty of combining pose data with displacements. Despite input standardization, the network appears to have difficulty processing these distinct feature types. We can observe in Figure~\ref{fig:displacement:error} that, for both architectures, the error increases monotonically in most cases.

The rest of the experiments show similar results with minor variations. Removing the cumulative sum term in the displacement loss leads to worse performance, particularly noticeable in Figure~\ref{fig:displacement:error}, where the error exhibits a consistent increase over time. Including both pressure and IMU information yields comparable results to using only IMU data, with the most significant differences observed in our dataset, particularly during idle poses like squatting. We assume that including pressure information may increase the risk of overfitting, while relying solely on IMU data improves generalization. However, this behavior may change with larger training datasets. Overall, our full approach, with or without pressure data, consistently achieves the best results.

\subsection{Analysis of Specific Motion Types}
\label{sec:experiments}
In this section, we examine \titleshort{} in diverse settings, starting with locomotion reconstruction using our dataset and extending our analysis to more challenging motions, such as dancing, walking on tiptoes, or \textit{in-the-wild} capture. We refer the reader to the companion video for additional qualitative results.

\begin{figure}
\centering
    \includegraphics[width=1\linewidth]{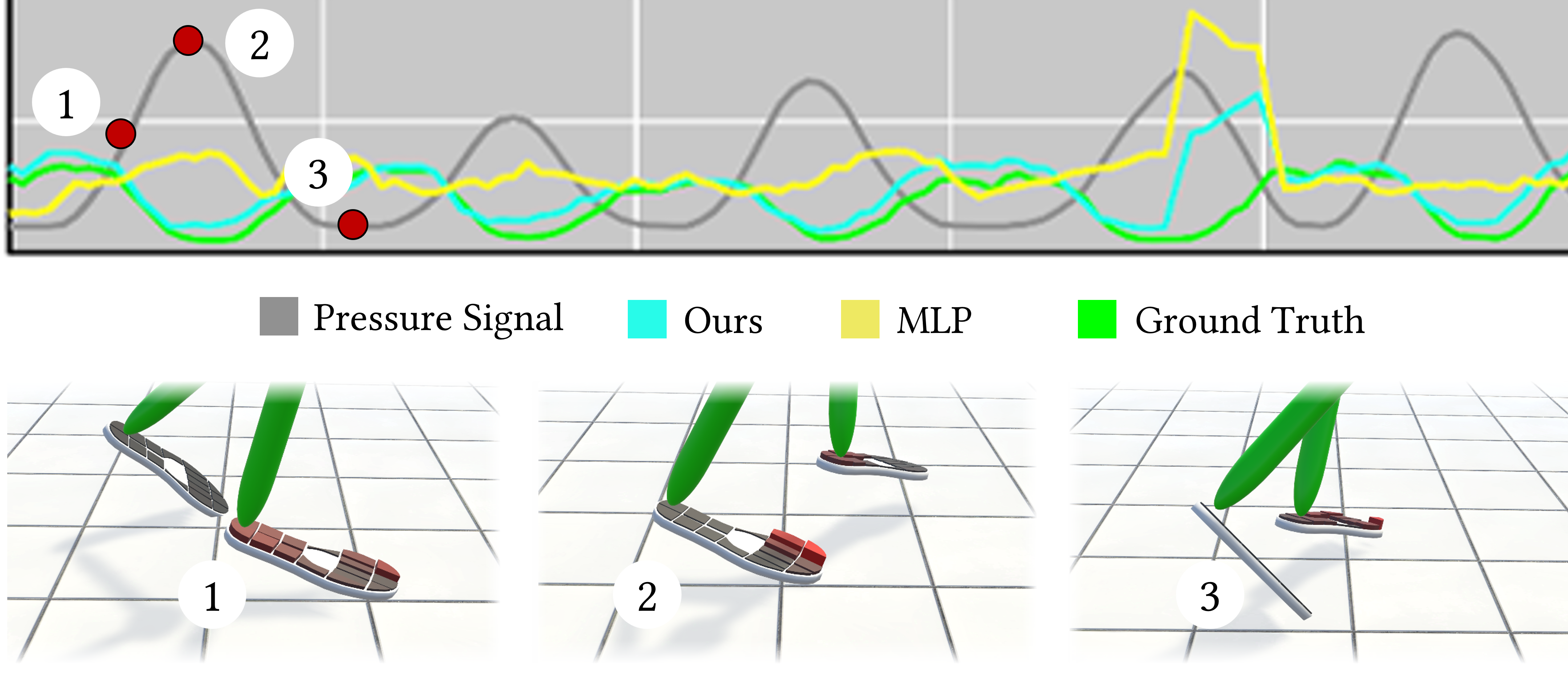}
    \caption{
    \textbf{Comparison of acceleration against pressure curves for a jogging sequence.} Our method (cyan) accurately captures jogging acceleration patterns, with increases aligning with pressure release, unlike the overly smooth \textbf{MLP} baseline (yellow). The \textbf{Transformer} baseline is excluded due to large fluctuations. The visualization includes left toes pressure (white) and ground truth acceleration (green).
    }
    \label{fig:experiments:pressure_acc_curve}
\end{figure}

\begin{figure}[htb]
     \centering
     \includegraphics[width=1\linewidth]{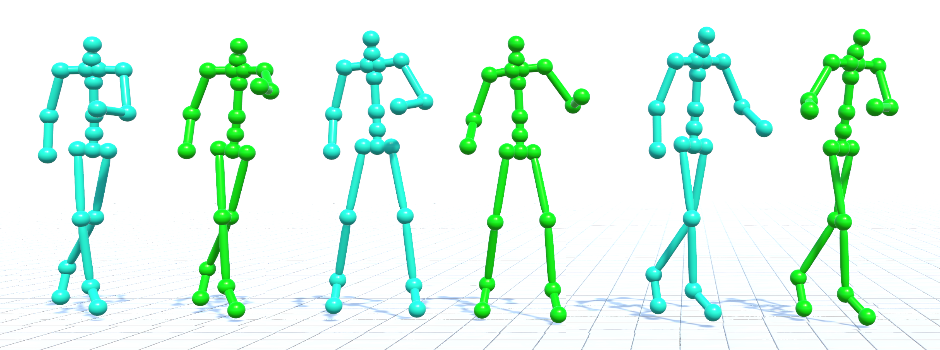}
     \caption{\textbf{Dancing animation reconstructed by Step2Motion}. Trained on dancing motion, it accurately reconstructs the lower-body movements and temporally coherent upper-body motion for unseen motion during training.}
     \label{fig:experiments:dancing}
 \end{figure}

\paragraph*{Locomotion.}
We demonstrate our method on different locomotion tasks. Sections~\ref{sec:accuracy:pose}~and~\ref{sec:accuracy:displacement} provide a full accuracy evaluation. Our method reconstructs from basic locomotion tasks like walking, as visualized in Figure~\ref{fig:teaser} (middle), to more complex movements, such as dancing, shown in Figure~\ref{fig:experiments:dancing}, thereby demonstrating its ability to predict accurate gait cycles and plausible full-body motion. Additional reconstructions for diverse tasks are shown in Figure~\ref{fig:experiments:diverse_motion}.

To further validate the relationship between foot pressure and the reconstructed motion, we examine the acceleration curves generated by our method and the baselines compared to the pressure data. Specifically, we selected the pressure sensor on the left insole's toe region. 
During the walking gait, increased exerted pressure by the toes signals the beginning of the swing phase. This pressure peaks during the toe-off phase, immediately before the increase in toe acceleration. Therefore, we expect to observe an increase in acceleration between the peak pressure and its subsequent decrease.
%
%
This relationship is demonstrated in Figure~\ref{fig:experiments:pressure_acc_curve}. A jogging animation is visualized, where our method closely follows the ground truth acceleration, exhibiting distinct peaks in acceleration that align with the release of pressure. In contrast, the \textbf{MLP} baseline produces a relatively flat acceleration curve, averaging the predicted acceleration throughout the motion sequence.

\paragraph*{In-the-wild capture.}
To assess the performance of our method in real outdoor settings, we conduct an \textit{in-the-wild} experiment, capturing insole data from a user performing various locomotion tasks while wearing the insole sensors. The motion is then reconstructed using our system trained on our dataset. In this case, linear accelerations from the IMU are transformed to a fixed world frame by integrating the angular rates.
Figure~\ref{fig:teaser} (right) shows a reconstructed locomotion sequence, which demonstrates \titleshort{}'s ability to recover plausible motion sequences in outdoor environments. 

\begin{figure}
    \centering
    \includegraphics[width=1\linewidth]{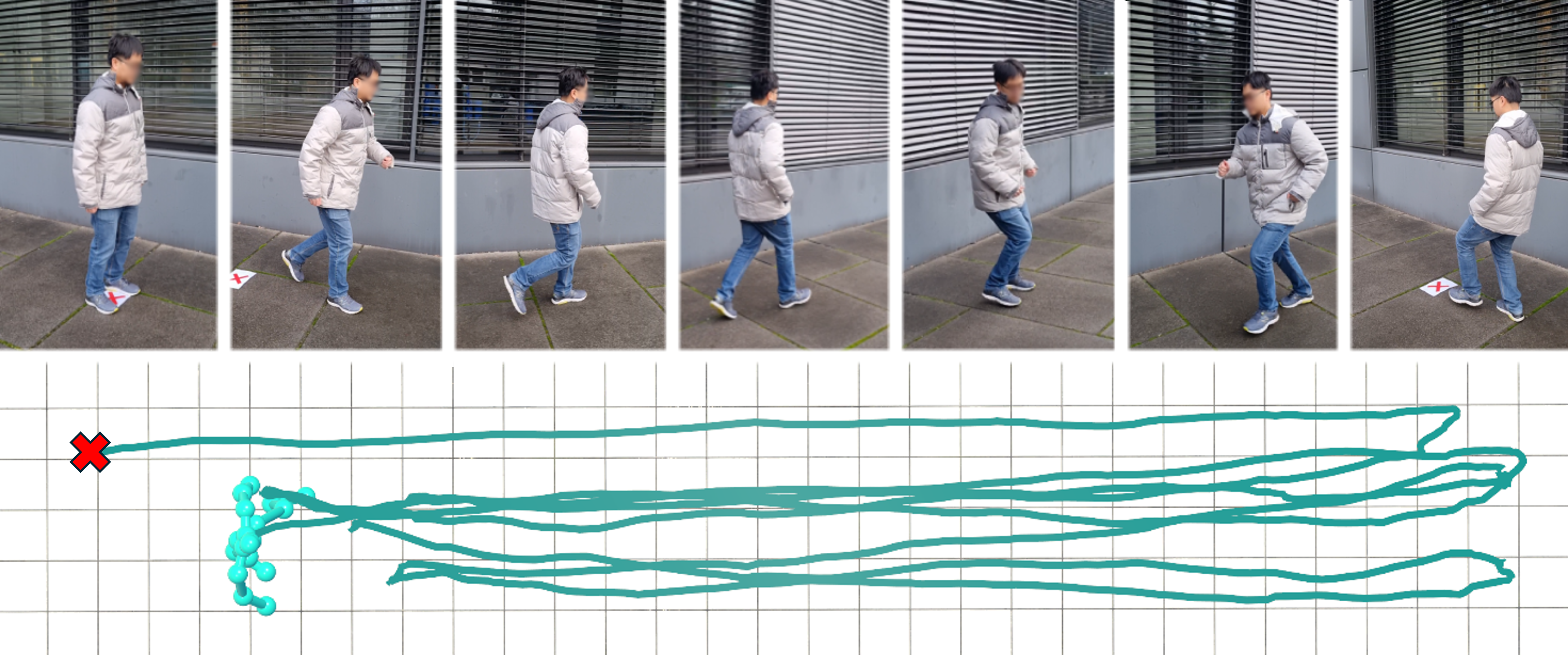}
    \caption{\textbf{Visualization of positional drift in \textit{in-the-wild} capture.} A user jogged repeatedly from the red cross marker to a point 7.5 meters away and back, completing four cycles. The total distance covered was 60 meters. The final drift, indicated by the offset between the red cross and the character's end position, is approximately 0.75 meters (1.25\,\% of the total distance). }
    \label{fig:wild:displacement}
\end{figure}

We also analyze the displacement accuracy for \textit{in-the-wild} capture. Figure~\ref{fig:wild:displacement} shows the root trajectory of a user repetitively jogging from a marked point to a location 7.5 meters away and back, completing four cycles for a total distance of 60 meters. The final drift of our full approach is approximately 0.75 meters ($1.25\,\%$ of the total distance). The displacement predictor baselines (see Section~\ref{sec:accuracy:displacement}) achieve considerably worse results: \textbf{Double Integration} (9.8 meters, $16.33\,\%$), \textbf{MLP} (3.85 meters, $6.41\,\%$), \textbf{Ours (only Pressure)} (no movement in the whole sequence), \textbf{Combined} (4.6 meters, $7.66\,\%$), \textbf{Ours w/o Cumsum Loss} (2.45 meters, $4.08\,\%$), and \textbf{Ours (Pressure+IMU)} (3.25 meters, $5.41\,\%$).

\paragraph*{Dancing.} 
To assess \titleshort{} in a more challenging reconstruction task, we trained our model to predict \textit{dancing} motions. Due to the scarcity of paired insole sensors and motion capture data, we collected and trained our system in a limited dataset of dance motions (approximately 15 minutes), focusing on a specific dance style. We train on 12 minutes of data and use 3 minutes for testing. Figure~\ref{fig:experiments:dancing} and the companion video showcase the results. \titleshort{} effectively reconstructs temporally coherent dance sequences, including natural arm movements that align with the lower-body motion. A quantitative evaluation of these dance data is provided in Section~\ref{sec:eval:insole}, analyzing the importance of utilizing both pressure and IMU information for accurate reconstruction. The dancing data used in this analysis will be released together with the \titleshort{} dataset.

\subsection{Influence of Multimodal Insole Data}
\label{sec:eval:insole}

Section~\ref{sec:accuracy:pose} focuses on analyzing pose reconstruction accuracy on a publicly available dataset, primarily evaluating locomotion and jumping animations. To further assess the potential of combining pressure and IMU data, we now analyze dance motions (see \textit{Dancing} in Section~\ref{sec:experiments}), which present a greater challenge and require both data modalities for meaningful pose reconstruction.

Table~\ref{tab:imu:pressure} presents the results of evaluating our full approach, \textbf{Ours}, against variations using only IMU data, \textbf{Ours (only IMU)}; only pressure data, \textbf{Ours (only Pressure)}; and replacing the insole multi-head cross-attention with a standard cross-attention module, \textbf{Ours w/o Insole MHA}. The latter effectively removes the specialized mechanism for incorporating insole sensor data into the diffusion process. Our results demonstrate that only the full approach successfully reconstructs dance motions. In the other three scenarios, during training, the validation loss stops decreasing in early epochs, indicating the model's inability to learn a relationship between the sensor readings and the target motion. This results in significantly worse performance when using only pressure or IMU information and when excluding our insole cross-attention.

\begin{table}[ht]
     \caption{\textbf{Accuracy evaluation of pose reconstruction on dancing motion.} This analysis demonstrates the efficacy of \titleshort{} in combining pressure and IMU information for reconstructing complex motions like dancing, which are challenging to reconstruct using IMU data alone. Errors are reported in cm and cm/s.}
     \centering
     \begin{tabular}{lccc}
     \toprule
         \multirow{2.5}{*}{Method} & \multicolumn{3}{c}{Dancing} \\
     \cmidrule(lr){2-4}
         & \emph{MPJPE} $\downarrow$ & $\emph{MPJPE}_{\text{Legs}}$ $\downarrow$ & $\emph{MPJVE}_{\text{Legs}}$ $\downarrow$ \\
     \midrule
         w/o Insole MHA & 31.6(25.5) & 34.4(16.2) & 307.1(193.4) \\
         Only Pressure & 25.7(22.1) & 26.1(14.5) & 296.9(246.3) \\
         Only IMU & 21.3(18.7) & 22.7(14.9) & 320.9(275.9)  \\
         Ours & \textbf{8.2(10.5)} & \textbf{6.9(6.6)} & \textbf{34.2(40.5)} \\
     \bottomrule
     \end{tabular}
     \label{tab:imu:pressure}
 \end{table}

\section{Limitations and Future Work}
\label{sec:limitations:futurework}

A main limitation of our method stems from the inherent drift associated with IMU sensors and the limited measurements on body parts far from the feet, i.e., head and arms, affecting the recovery of certain motions like squats.
This primarily affects the accuracy of root motion and arm pose prediction, but can also influence the predicted global orientation when significant drift occurs in angular rate measurements. 
Since insole data primarily dictates lower-body kinematics, upper-body reconstruction is inherently probabilistic. Our model generates \textit{plausible} motion based on learned correlations (e.g., arm swing), rather than deterministic tracking.

We also observe foot sliding, which is common in generative approaches. While our method implicitly learns contact from pressure, it does not explicitly enforce hard constraints. Future work could leverage the physical nature of pressure data to enforce foot-lock constraints during a post-optimization stage. 

While using only the IMUs helps prevent overfitting in displacement prediction, more motion data paired with insole readings (ideally incorporating other sensor modalities like egocentric cameras or additional IMUs) could help mitigate this issue.
Additionally, generating synthetic insole readings from existing motion sequences could substantially increase the amount of training data, which would require an accurate physics-based model for foot-ground interactions and soft-tissue deformation.
Another potential solution involves using motion priors trained on large motion capture databases and fine-tuning them for insole sensors. However, the domain gap is significant, and large research datasets lack sufficient diversity and quality. We found that training a specialized architecture from scratch yielded better alignment with the insoles signal. We leave further exploration for future work, as our initial tests suggested that significant work would be required to successfully use pre-trained models.

Integrating the pressure readings in the attention mechanism is still open for further study. Our approach divides pressure into two key regions: toes and heel, motivated by their significant roles in the push-off and heel strike phases during gait.
This could be improved by adopting a finer sole pattern to increase accuracy.

\section{Conclusions}
\label{sec:conclusions}

This work represents a significant first step towards general locomotion reconstruction only from insole sensors. To the best of our knowledge, \titleshort{} is the first approach to tackle this challenging problem, demonstrating that insole readings can be excellent, feature-rich human motion descriptors.

Our approach leverages the power of diffusion models, combined with a dedicated displacement prediction network and a carefully designed cross-attention mechanism, to effectively capture the complex relationship between human movement and the different sensors included in the insoles. Through extensive evaluations on a public database and our dataset, we demonstrated \titleshort{}'s ability to accurately reconstruct lower-body animation while synthesizing plausible upper-body movement.

This work opens exciting possibilities for future research in motion capture and animation. The ability to reconstruct detailed human motion using only insole sensors could greatly benefit applications in sports analysis, rehabilitation, and entertainment, enabling more accessible and versatile motion capture in unconstrained environments.

\section{Acknowledgements}
This work has received funding from MCIN/AEI/10.13039/ 501100011033/FEDER, UE (Spain) in the framework of the project PID2021-122136OB-C21, and with the support of the Department of Research and Universities of the Government of Catalonia (2021 SGR 01035). Jose Luis Ponton was also funded by the Spanish Ministry of Universities (FPU21/01927 and EST24/00555).


\printbibliography   


\end{document}